# Delusions of Success

Invited comment for *Harvard Business Review*

By Bent Flyvbjerg

Full reference: Bent Flyvbjerg, 2003, "Delusions of Success: Comment on Dan Lovallo and Daniel Kahneman," *Harvard Business Review*, December, pp. 121-122.

Link to comment:
https://dl.dropboxusercontent.com/u/14316186/HBR0312ASPUBLISHED%20copy.pdf

Dan Lovallo and Daniel Kahneman must be commended for their clear identification of causes and cures to the planning fallacy in "Delusions of Success: How Optimism Undermines Executives' Decisions" (July 2003). Their look at overoptimism, anchoring, competitor neglect, and the outside view in forecasting is highly useful to executives and forecasters. However, Lovallo and Kahneman underrate one source of bias in forecasting—the deliberate "cooking" of forecasts to get ventures started. My colleagues and I call this the Machiavelli factor.

The authors touch upon this issue in their mention of the organizational pressures forecasters face to exaggerate potential business results. But adjusting forecasts because of such pressures can hardly be called optimism or a fallacy; deliberate deception is a more accurate term. Consequently, Lovallo and Kahneman's analysis of the planning fallacy seems valid mainly where political pressures are insignificant. When organizational pressures are significant, both the causes and cures for rosy forecasts will be different from those described by the authors.

In our study of bias in cost and demand forecasting in capital-investment transport projects, my colleagues and I found strong evidence of heavy political pressures on executives to make rosy forecasts and of minor penalties for having made such forecasts. Indeed, during the 70 years covered by our study, forecasters consistently made errors of the same size and frequency, resulting in the same cost overruns and demand failures over and over. Urban rail investments, for instance, had average cost overruns of 45% in constant prices, while actual patronage on average was 50% lower than forecasted. Comparing our data to data for other types of investments, we found the same pattern. Forecasts for public works projects are not fundamentally different from those for, say, private dot-com IPOs, in this respect.

Overoptimism, as depicted by Lovallo and Kahneman, would be an important and credible explanation of this phenomenon if estimates were produced by inexperienced forecasters. But given the fact that humans can, and do, learn from experience, it seems unlikely that forecasters would continue to make the same mistakes decade after decade. It seems even more unlikely that whole professions of specialized forecasters, often called in from outside the organization as Lovallo and Kahneman recommend, would collectively be subject to such a bias and would not learn over time. Such learning would help to reduce, if not eliminate, overoptimism, and then cost and demand estimates would become more accurate over time. But our data clearly shows this has not happened.

We tested our data against four different explanations for bias in forecasting—technical, economic, psychological, and political—and found that political explanations best fit the data. This has been confirmed by interviews with forecasters.

If executives deliberately cook their estimates of costs and benefits because of organizational pressure to do so, it will be harder to arrive at more realistic forecasts than if executives are the involuntary victims of delusional optimism. In the latter case, which is the one Lovallo and Kahneman describe, executives will be motivated to embrace the outside view and get forecasts right. In the former case, which appears to be more common, there will be little incentive to accept the outside view.

**Bent Flyvbjerg**
*Professor of Planning*
*Department of Development and Planning*
*Aalborg University*

*Denmark*